\documentstyle[preprint,prb,aps]{revtex}
\begin{document}
\draft
\def\gappeq{\mathrel{\rlap {\raise.5ex\hbox{$>$}}
{\lower.5ex\hbox{$\sim$}}}}
 
\def\lappeq{\mathrel{\rlap{\raise.5ex\hbox{$<$}}
{\lower.5ex\hbox{$\sim$}}}}
 
\def \gsim{\lower.8ex\hbox{$\sim$}\kern-.75em\raise.45ex\hbox{$>$}\;}
\def \lsim{\lower.8ex\hbox{$\sim$}\kern-.8em\raise.45ex\hbox{$<$}\;}
 
                      \def\ltsima{$\; \buildrel < \over \sim \;$}
                      \def\simlt{\lower.5ex\hbox{\ltsima}}
                      \def\rtsima{$\; \buildrel > \over \sim \;$}
                      \def\simrt{\lower.5ex\hbox{\rtsima}}

   \title{Model of the Stochastic Vacuum and QCD Parameters\\}
 
\author{Erasmo Ferreira}
\address{
Instituto de F\'\i sica, Universidade Federal do Rio de Janeiro\\
 Rio de Janeiro 21945-970, RJ, Brazil\\}
 
\author{Fl\'avio Pereira}
\address{ DAGE, Observat\'orio Nacional, CNPq       \\
 Rio de Janeiro 20921-400, RJ, Brazil\\}

 
  \maketitle
 
\begin{abstract}
     Accounting for the two independent correlation functions of the 
QCD vacuum, we improve the simple and consistent description
given by  the model of the stochastic vacuum to the high-energy
pp and $\bar {\rm p}{\rm p}$ data, with a new determination of  
parameters of non-perturbative QCD. The increase of the hadronic radii 
with the energy accounts for the energy  dependence of the observables.

   \end{abstract}
\bigskip
 PACS Numbers~:~12.38 Lg , 13.85 -t, 13.85 Dz , 13.85 Lg~.

\newpage

{\bf 1. Profile Function for Hadron-Hadron Scattering }
 
 In the effort to explain the properties of soft high-energy 
scattering in a non-pertubative QCD framework, 
 we reconsider  the application \cite{RRG3} of the model of the
stochastic vacuum of non-perturbative QCD to hadron-hadron scattering. This 
approach combines the parameters describing properties of the QCD field 
(gluon condensate, correlation length) with those describing the colourless 
hadrons. The model presents the characteristic 
features of the pomeron exchange mechanism of Regge phenomenology, where 
vacuum quantum numbers are exchanged between hadronic structures. For all
hadronic systems the total cross-sections increase with the energy 
\cite{RRB2,RRB2A} somewhat between $s^{0.0808}$ and $\log ^2(s/s_0)$
within the present experimental range, and this is explained in terms of 
the energy dependence of the effective hadronic radii. 

Several models relate the total high-energy cross-sections and slope 
parameters to the hadronic radii \cite{RRB5,RRB5A}. 
This is a characteristic feature also of the model of the
stochastic vacuum which gives specific predictions for the size dependence
of the observables for different hadronic
systems  \cite{RRG3}. 
The knowledge of the hadronic structures required for the
description of the soft high-energy data does not go beyond the
information on their sizes, the simplest transverse
wave-function giving all information required for the determination of the 
observables. 

     The model of the stochastic vacuum deals successfully with 
non-perturbative effects both in low-energy hadron physics \cite{RRC1}
and in soft high-energy scattering \cite{RRG3}  . 
The treatment of scattering is based on the concept of loop-loop 
interaction, 
which allows a gauge-independent formulation for the amplitudes. 
The loops, formed by the quark and antiquark light-like paths in a 
moving  hadron, have their contributions  added incoherently, with 
their sizes weighed by transverse hadronic wave-functions.

The principles and methods used for the evaluation of the 
observables of high-energy scattering in the model of the stochastic 
vacuum have been fully described before \cite{RRG3}. 
In the  present work we present the results of a more complete calculation
in which both Abelian and non-Abelian contributions to the field 
correlator are taken into account. 
The dynamics is generated by correlations in the QCD vacuum field, 
according to the pioneering work of Nachtmann \cite{RRD7}.
With the assumption that the correlator is independent of the
reference point of the gauge field, its most general form \cite{RRC1} 
contains two tensor structures, each with a correlation function, with a 
weigth parameter $\kappa$ between them. 

    To write the expression for the profile function for hadron-hadron 
scattering, we introduce the notation $\vec R(I,J)$, where the first index
(I=1,2) specifies the loop, and the second specifies the particular
quark or antiquark (J=1 or 2) in that loop.
      Fig.1 shows a projection of the collision of two mesons 
on the transverse scattering plane.
                         The vectors $\vec Q(K,L)$ in the transverse
plane connect the reference point $C$  to the
positions of the quarks and antiquarks of the loops 1 and 2. The quantity
$\psi(K,L)$ is the angle between $\vec Q(1,K)$ and $\vec Q(2,L)$.

The eikonal function of the loop-loop amplitude is written 
\begin{eqnarray}
  \widetilde \chi ( \vec b,\vec {R}(1,1),\vec {R}(2,1))= \kappa \bigg[
          -\cos \psi (1,1)~I[Q(1,1),Q(2,1),\psi (1,1)] ~~ \nonumber \\
          -\cos \psi (2,2)~I[Q(1,2),Q(2,2),\psi (2,2)] ~~ \nonumber \\
          +\cos \psi (1,2)~I[Q(1,1),Q(2,2),\psi (1,2)] ~~ \nonumber \\
          +\cos \psi (2,1)~I[Q(1,2),Q(2,1),\psi (2,1)] ~ \bigg] \nonumber \\
     +(1-\kappa) \bigg[ -W[Q(1,1),Q(2,1),\psi (1,1)]
          -W[Q(1,2),Q(2,2),\psi (2,2)] ~~ \nonumber \\
          +W[Q(1,1),Q(2,2),\psi (1,2)]
          +W[Q(1,2),Q(2,1),\psi (2,1)] ~ \bigg] ~.
 \label{4A6}
\end{eqnarray}
     The quantities {\it {I}} which represent the non-Abelian
contributions  are given by integrations along the dashed lines of the figure
 
\begin{eqnarray}
  I[Q(1,K),Q(2,L),\psi (K,L)] = {\frac {32}{9\pi}}
                         \bigg(\frac{3\pi}{8}\bigg)^2
\quad\quad\quad\quad\quad\quad\quad\quad\quad \quad \nonumber \\
\quad\quad  \times \{  Q(1,K)
   \int_{0}^{Q(2,L)} [Q(1,K)^2+x^2-2 x Q(1,K) \cos{\psi(K,L)} ]
                                     \nonumber  \\
K_{2}\bigg[\frac{3\pi}{8}\sqrt{Q(1,K)^2+x^2
    -2 x Q(1,K) \cos{\psi(K,L)}}\bigg]
                                ~dx ~~~                 \nonumber \\
                                                        \nonumber \\
    + Q(2,L)
   \int _{0}^{Q(1,K)}  [Q(2,L)^2+x^2-2 x Q(2,L) \cos{\psi(K,L)} ]
                                     \nonumber  \\
K_{2}\bigg[\frac{3\pi}{8}\sqrt{Q(2,L)^2+x^2
    -2 x Q(2,L) \cos{\psi(K,L)}}\bigg]
                                ~dx               \}~,
\label{4A7}
\end{eqnarray}
with  $Q(K,L)=\vert\vec Q(K,L)\vert$. The quantities W, which come from the
non-confining part of the correlator, are given by
 
\begin{eqnarray}
  W[Q(1,K),Q(2,L),\psi (K,L)] = \frac {32}{9\pi}~ 2 ~
                         \frac{3\pi}{8} \nonumber \\
  \times 
   [Q(1,K)^2+Q(2,L)^2-2 Q(1,K)Q(1,L) \cos{\psi(K,L)} ]^{3/2}
                                     \nonumber  \\
K_{3}\bigg[\frac{3\pi}{8}\sqrt{Q(1,K)^2+Q(2,L)^2
    -2 Q(1,K) Q(2,L) \cos{\psi(K,L)}}\bigg] ~.
 \label{4A7B}
\end{eqnarray}
   From the eikonal function $\widetilde \chi$ we form the loop-loop
amplitude $ \widetilde J_{L_1 L_2}(\vec b, \vec R_1, \vec R_2)$ through
\begin{eqnarray}
    \widetilde J_{LL'}(\vec b, \vec R_1, \vec R_2) \equiv
\frac{1}{\big[ \langle g^2 FF \rangle\big]^2}
 J_{LL'}(\vec b, \vec R_1, \vec R_2)=
-\frac{[ \widetilde \chi(\vec b, \vec R_1, \vec R_2)  ]^2 }
{144 \cdot 576}~,
\label{4A5}
\end{eqnarray}
 where $\vec R_1$ and $\vec R_2$ are shorthand
notations for $\vec R(1,1)$ and $\vec R(2,1)$ respectively.

   The hadron-hadron amplitude  is constructed from the loop-loop 
amplitude using a simple quark model for the hadrons, 
by smearing over the values of
$\vec R_1$ and $\vec R_2$  with transverse wave-functions
$\psi(\vec R)$.  
Taking into account the results of the previous analysis of different
hadronic systems  \cite{RRG3}, in the present calculation we
only consider for the proton a diquark structure , where the proton
is described as a meson, in which the diquark replaces the antiquark.
Thus these expressions apply equally well to meson-meson, meson-baryon 
and baryon-baryon scattering. 
  
           For the hadron transverse wave-function we make the 
simple ansatz
\begin{equation}
     \psi _{H} (R)=  \sqrt {2/\pi}\frac {1}{S_H} \exp{(-R^{2}/S_H^{2})}~,
\label{4A12}
\end{equation}
     where $S_H$  is a parameter for the hadron size.
    We then write the reduced profile function of the eikonal amplitude
 \begin{equation}
 \widehat J_{H_1 H_2}(\vec b,S_1,S_2)=
       \int d^{2}\vec R_{1}\int d^{2}\vec R_{2}~
  \widetilde J_{L_1 L_2}(\vec b,\vec R_1,\vec R_2)~
           {\vert\psi_1(\vec R_1)\vert}^2
           {\vert\psi_2(\vec R_2)\vert}^2~,
 \label{4A13}
 \end{equation}
    which is a dimensionless quantity.
       The dimensionless scattering amplitude is given by
\begin{equation}
 T_{H_1 H_2} = i s [\langle g^2 FF\rangle a^4]^2 a^{2}\int d^2 \vec b~
          \exp{(i \vec q\cdot \vec b )}~\widehat J_{H_1 H_2}
 (\vec b,S_1,S_2) ~,
 \label{4A14}
 \end{equation}
  where the impact parameter vector  $\vec b$  and the hadron sizes
  $S_1$ , $S_2$ are in units of the
     correlation length $a$, and $ \vec q $ is the momentum
     transfer projected on the transverse plane, in units of $1/a$,
     so that the momentum transfer squared is $t=-|\vec q|^2/a^2$. 
For convenience we have explicitly factorized the 
dimensionless combination  $\langle g^2 FF\rangle a^4 ~.$ 
        The normalization  for $T_{H_1 H_2}$ is such that the total
and differential cross-sections are given by 
\begin{eqnarray}
 \sigma^T  = \frac{1}{s}~\hbox{\rm Im}~T_{H_1 H_2} ~~~,~~~
    \frac {d\sigma^{e\ell}}{dt} =
             \frac {1}{16\pi s^2}~\vert T_{H_1 H_2}\vert^2  ~.
\label{4A15}
\end{eqnarray}
 
    For short, from now on we write $J(b)$ or $J(b/a)$  to represent
$\widehat J(\vec b,S_1,S_2)$.
        To write convenient expressions for the observables, we 
 define the dimensionless moments of the profile function 
(as before, with $b$ in units of the correlation length $a$)
\begin{equation}
              I_k = \int d^2\vec b~b^k~ J(b) ~~,~k=0, 1, 2, ...
\label{4A18}
\end{equation}
which depend only on $ S_1/a$, $ S_2/a$, and the Fourier-Bessel transform
\begin{equation}
              I(t) = \int d^2\vec b~ J_0 (b a \sqrt{|t|})~ J(b) ~,
\label{4A19}
\end{equation}
where $J_0(b a \sqrt{|t|})$ is the zeroth--order Bessel function.
Then $$T_{H_1 H_2} = i s[\langle g^2 FF\rangle a^4]^2 a^{2} I(t)~. $$
  Since  $J(b)$ is real, the total cross section $\sigma^T$ and the
slope parameter B (slope at $t=0$) are given by
\begin{eqnarray}
            \sigma^T= I_0~[\langle g^2 FF \rangle a^4]^2 a^{2} ~~,~~
   B = \frac {d}{dt} \biggl( \ln \frac {d\sigma^{e\ell}}{dt} \biggr)
   \bigg\vert_{t=0} ~=\frac{1}{2}~ \frac{I_2}{I_0} ~a^2 ~ \equiv K a^2~.
\label {4A21}
\end{eqnarray}

   The model conveniently factorizes the  QCD strength and length scale in
the expressions for the observables, and the correlation length appears as
the natural unit of length for the geometric aspects of the interaction.
These aspects are contained in the quantities $I_0(S_1/a,S_2/a)$ and
$I_2(S_1/a,S_2/a)$, which depend on the hadronic structures and on the shapes
and relative weights (parameter $ \kappa $) of the two correlation 
functions. It has been shown before \cite{RRG3} that for the case $\kappa=1$
(pure confining correlator)
the two moments have simple form as functions of $S/a$, and  
in the present work we have obtained similarly simple 
expressions for any value of $ \kappa $. We concentrate on the range about 
$\kappa=3/4$, which is indicated by lattice results. 

It is important that the high-energy observables $  \sigma^T$ and $B$
require only the two low moments $I_0$ , $I_2$ of the profile functions.
    We observe that in the lowest order of the correlator expansion the  
slope parameter $B$ does not depend on the value of the gluon
condensate  $ \langle g^2 FF \rangle $ and, once the proton radius $S$ 
is known, it may give a direct determination of the correlation length.

\bigskip  

{\bf 2. pp and $\bar {\rm p}$p systems and QCD Parameters}

 In order to have a wide range of data to extract reliable 
information on QCD parameters, we concentrate here  on the 
 pp and $\bar {\rm p}{\rm p}$ systems, for which $S_1=S_2=S$. 
    The curves for  $I_0= \sigma^T /~[\langle g^2 FF \rangle^2 a^{10}] $~ 
and $ K= B/a^2 $ can be parametrized as functions of $S/a$ with simple 
powers, with good accuracy. The convenient expressions are  
\begin{eqnarray}
I_0= \alpha \bigg(\frac{S}{a}\bigg)^{\beta}~~~,~~~ 
          K=\eta +\gamma \bigg(\frac{S}{a}\bigg)^{\delta}~.
\label{4A27}
\end{eqnarray}
The numerical values for the parameters are given in table 1.
 
\bigskip

The parametrizations for $\sigma^T$ and $B$ are  very convenient
for comparison of the  results of the model of the stochastic vacuum with
experiment.  In the present analysis we take into account all
 available data on $\sigma^T$ and B  in 
pp and $\bar {\rm p}$p scattering , which
consist mainly \cite{RRDAT1,RRA2} of ISR (CERN) 
measurements at energies ranging from $\sqrt{s}=23$ GeV to 
$\sqrt{s}=63$ GeV, of the $\sqrt{s}=541 - 546$  GeV  measurements
in CERN SPS and in Fermilab, and of the $\sqrt{s}=1800$ GeV
information from the E-710 Fermilab experiment. Besides these, there is 
a measurement \cite{RRDAT2} of $\sigma^{T}=65.3\pm2.3$ mb
at  $\sqrt{s}=900$ GeV and there are the measurements of 
$\sigma^{T}=80.6\pm 2.3$ and $B=17.0\pm 0.25$ GeV$^{-2}$ 
in  Fermilab CDF \cite{RRDAT3} at $\sqrt{s}=1800$ GeV
which seem discrepant with the E-710 experiment at the same energy.
 A measurement by Burq et al.\cite{RRA3} at $\sqrt{s}=19$ GeV, which was 
taken as input in a previous calculation   \cite{RRG3}, seems to disagree 
with the ISR data, presenting a too high slope
$B=12.47\pm 0.10$ GeV$^{-2}$. 

In fig.2 we plot $\sigma^T$ and $B$ against each other.
At the ISR energies we use $\sigma^T_{\rm pom}=(21.70~{\rm mb}) s^{0.0808}$
and the pp slope parameter
as representative of the non-perturbative contributions which are the
concern of the model of the stochastic vacuum.  
At the highest energies (541-1800 GeV) the process is essentially 
non-perturbative and dominated by pomeron exchange. The relation 
between the two observables is fitted with the form
\begin{equation}
B=B_\Delta + C_\Delta  (\sigma^T)^\Delta~.
\label{1A1}
\end{equation} 
This form, which is represented by 
the solid line in the figure, is suggested by the results of 
the calculations with the model of the stochastic vacuum \cite{RRG3}, where 
an interpretation for the meaning of the parameters is given in terms of QCD 
and hadronic quantities. 

\bigskip

The parameters in the
model that are fundamentally related to QCD are the weigth $\kappa$, the gluon
condensate $\langle g^2 FF\rangle$ and the correlation length $a$. The
hadronic extension parameter  $S_H$ accounts for the energy dependence 
of the observables. We here show how these quantities  can be 
evaluated using only high-energy scattering data.
 
     The proton radius can be eliminated from
 eqs(\ref{4A21}) and  (\ref{4A27}), and we obtain 
a relation between the observables $\sigma^T$ and $B$ at a given energy  
\begin{equation}
(B - \eta a^2)= \frac{a^2}{(<g^2 FF>a^4)^{2\delta/\beta}}
           \frac{\gamma}{\alpha^{\delta/\beta}}
\bigg(\frac{\sigma^T_{\rm pom}}{a^2}\bigg)^{\delta/\beta}~.
\label{5A29}
\end{equation}
 This form is the same as given by eq.(\ref{1A1}),  
with an obvious correspondence of parameters.  
We first remark that the exponent 
$\Delta=\delta/\beta $ does not depend on QCD quantities and is almost
constant (equal to about 0.75) in the range of values of $\kappa$ that are 
obtained in lattice calculations ($\kappa\approx 3/4$). 
Then we fix $\Delta=\delta/\beta=0.75$ and 
are left with only two free quantities in both energy independent
relations  (\ref{1A1})and  (\ref{5A29}), and they  can be determined using
as input the two clean experimental points for $\sigma^T$ and B at 541 and 
1800 GeV. The results for the parameters are nearly the same as obtained
in a free fitting of all points. With $\sigma^T$ in mb we have 
 \begin{equation}
 B_\Delta=\eta a^2=5.38~{\rm GeV}^{-2}=0.210~{\rm fm}^{2}~~~,~~~ 
            C_\Delta= 0.458~{\rm GeV}^{-2}~.
    \label{5A1D}
 \end{equation}

Of course these results are subject to uncertainties. We have adopted an 
ansatz for the correlation function, which is arbitrary (although 
numerically it could not have very different shape). There is some 
uncertainty also 
in the determination of the parameters $\alpha, \beta \dots $ 
representing the final results of the numerical calculation. 
On the other hand, the model gives a rather unique prediction for
$\Delta=\delta/\beta=0.75$ and this value is well sustained by the data 
as shown in fig.2 . 

We have made calculations with several values of $\kappa$, but to be 
specific, we borrow from lattice calculation the value $\kappa=3/4$,
and then use as parameter values the numbers shown in table 1. Taking into 
account the experimental error bars in the input data at 541 and 1800 GeV, 
we obtain
\begin{equation}
\kappa=\frac{3}{4},~a=0.32\pm 0.01~ {\rm fm}~,
    ~<g^2 FF>a^4 = 18.7\pm 0.4~,<g^2 FF>=2.7\pm 0.1~{\rm GeV}^4~.
\label{5A35A}
\end{equation}
   With the value $\kappa=33/40$ obtained in more recent lattice results
\cite{RRF2} the central values change only slightly to 
\begin{equation}
\kappa=33/40~,~a=0.33~ {\rm fm}~,
            ~<g^2 FF>a^4 = 19.2~,~ <g^2 FF>=2.6~{\rm GeV}^4~.
\label{5A35B}
\end{equation}

 The results of the pure $SU(3)$  lattice gauge calculation \cite{RRF1} for 
the correlator $\langle F^C_{\mu\nu} (x,0)\ F^D_{\rho\sigma}(0,0)\rangle_A$ 
have been fitted \cite{RRG3} with the same correlation function 
used in the present work. The correlation
between the values of  $<g^2 FF>$ and $a$ that was then obtained
can be well represented by the empirical expressions
\begin{equation}
 \Lambda_L=\frac{1.1122}{a^{1.310}}~~,~~<g^2 FF>=\frac{0.01813}{a^{4.656}}~~,  
~~ <g^2 FF> a^4 = 0.0172 \sqrt{\Lambda_L}~,
\label{5A38}
\end{equation}
with the lattice parameter $\Lambda_L$ in MeV, $a$ in fm, and $< g^2 FF>$ in 
GeV$^4$. This correlation is displayed in fig.3, for values of $\Lambda_L$ 
in the usual range.  The point representing our results in eq.(\ref{5A35A}) 
is marked in the same figure. The dashed line represents the relation with 
the string tension  obtained in the application of the model of 
the stochastic vacuum \cite{RRC1} to hadron spectroscopy; for our form of 
correlator, this relation is \cite{RRG3} 
\begin{equation}
\kappa\langle g^2FF\rangle=\frac{81 \pi}{8 a^2} \rho ~.
\label{5A8}
\end{equation}

\bigskip
 
 As can we may see in the figure, the constraints from these three
independent sources of information are simultaneously
satisfied, building a consistent
picture of soft high-energy $pp$ and $\bar p p$ scattering. 
 The (pure gauge) gluon condensate is well compatible with the expected 
value.  The lattice parameter $\Lambda_L$ and the
string tension $\rho$ are also in their acceptable ranges.
As we describe below, the resulting proton size parameter $S_p$ takes values 
quite close to the electromagnetic radius \cite{RRI1}.
 
In this model the increase of the observables with the energy is due to 
a slow energy dependence of the hadronic radii.
An explicit relation is obtained if
we bring into eqs.(\ref{4A21}) and (\ref{4A27}) a parametrization for the
energy dependence of the total cross-sections, such as the Donnachie-Landshoff
\cite {RRB2} form. In this case we obtain for the proton radius
\begin{equation}
S_p(s)= a\frac{1}{\alpha^{1/\beta}}
 \frac{1}{(\langle g^2 FF\rangle a^4)^{2/\beta}} 
    \bigg(\frac{21.7~{\rm mb}}{a^2}\bigg)^{1/\beta}
         ~s^{0.0808/\beta}~.
\label{5A19A}
\end{equation}
The energy dependence, given by a power $0.0808/\beta$ of s is very
slow, and the values obtained for $S_p$ are in the region of the proton 
electromagnetic radius  \cite{RRI1}, which is $R_p=0.862\pm 0.012$ fm. 
However the Donnachie-Landshoff parametrization for the total 
cross-sections is very convenient to identify the pomeron contributions
 at the ISR energies, but does not give the best representation
of the cross-sections at higher energies \cite{RRB2A}. Using    
eqs.(\ref{4A21}) and (\ref{4A27}) and directly the data at 541 and 1800
GeV, we obtain the values for the proton radius that are shown 
by the small squares in fig.4,
where a $\log_e$ scale is used for $\sqrt{s}$. It is remarkable that we have
an almost linear dependence, which can be represented by 
\begin{equation}
S_p(s)= 0.671 + 0.057 \log \sqrt{s}~~({\rm fm})~,
\label{5A19B}
\end{equation}
with $\sqrt{s}$ in GeV.  With this form for the radius, which is shown in 
dashed line in fig.4, the cross-sections
evaluated at very high energies rise with a term $\log^\beta{\sqrt{s}}$, and
are smaller than predicted by the power dependence of Donnachie-Landshoff. 
However, since $\beta\approx 2.8$, they still violate the bound
$\log^2{\sqrt{s}}$. This may be corrected using a power $2/\beta$ instead of
1 in the parametrization for $S_p(s)$, and we then obtain 
\begin{equation}
S_p(s)= 0.572 + 0.123~ [\log \sqrt{s}]^{0.72}~~({\rm fm})~.
\label{5A19C}
\end{equation}
This form is shown in solid line in fig.(4). Clearly it gives a good
representation for the existing data. At 14 TeV, which is the expected 
energy in the future LHC experiments, we obtain 
$S_p(14~TeV)=1.19~{\rm fm}=1.38~R_p=3.7a$~ and the 
model predictions for the observables are $\sigma^T$=92 mb and 
B=19.6 GeV$^{-2}$. The dashed line representing eq.(\ref{5A19B}) leads 
at the same LHC energy to $\sigma^T$=97 mb and 
B=20.1 GeV$^{-2}$, while the Donnachie-Landshoff formula leads to a 
higher value $\sigma^T$=101.5 mb. A recent fit of all data \cite{RRB2A}
gives a power $2.25\pm 0.35$  in the logarithm, and predicts 
$\sigma^T=112\pm 13$ mb at 14 GeV.  

\bigskip

  The non-perturbative QCD contributions to soft high-energy scattering are 
expected to be dominant in the forward direction, thus determining the 
total cross-section (through the optical theorem) and the forward slope 
parameter. The model, as it is presented in this paper, leads to a
negative curvature for the slope B(t), which decreases as $|t|$ increases.
The data however shows an almost zero curvature of the peak,
so that above some value of $|t|$ the model leads to too high values of
the differential cross-section. Rather small changes in the form of the 
profile function J(b) that enters in the expression for the scattering 
amplitude may modify this behaviour.
These changes can be made phenomenologically, introducing a form
factor \cite{RRL1}. However, in the present work we wish to keep the
fundamental characteristics of the model, which is that of a pure QCD based
calculation, with a unique set of quantities governing all systems at
all energies.

\bigskip
{\bf  3. Conclusions }

 Extending the previous calculation of high-energy observables, now 
including more data in the analysis, and taking into account the two 
independent correlators of the QCD gluon field, we give 
a unified and consistent description 
of all data on total cross-section and slope
parameter for the pp and $\bar {\rm p}$p systems, from 
$\sqrt{s}\approx$20 to 1800 GeV. The non-perturbative QCD parameters 
determining the observables, the gluon condensate and the correlation 
length of the vacuum field fluctuations, are determined. The third 
quantity entering the calculations is the transverse hadron size, which 
has a magnitude close to the electromagnetic radius, and whose 
slow variation accounts for the energy dependence  of the observables. 

The model allows a very convenient factorization between the QCD and 
hadronic sectors. The relation between $\sigma^T$ and B 
obtained by elimination of the hadron size parameter 
agrees very well with experiment. Starting from two experimental 
energies as input (541 and 1800 GeV), this expression allows a 
determination of the correlation length and gluon 
condensate from high-energy data alone and gives good prediction 
of the remaining data. The results obtained are in 
good agreement with the correlations between the two QCD parameters
obtained independently in lattice calculations and in the application of the
stochastic vacuum model to hadronic spectroscopy.

 The present calculation is restricted to the lowest order
non-vanishing contribution in the expansion of the exponential with 
functional integrations, and we may conclude from our results that this 
is justified for the evaluations of total cross-section and slope 
parameter. The resulting amplitude is purely imaginary, and the 
$\rho$-parameter (the ratio of the real to the imaginary parts of the
elastic scattering amplitude) can only be described if we go one
further order in the contributions to the correlator. 
 
It is interesting to compare our results with Regge phenomenology, where
the relation between the observables is given by 
$  \sigma^T_{\rm Regge} = \sigma^T_0~ e^{{0.1616} (B-B_0)}~$,
obtained from a Regge amplitude using the slope of the pomeron trajectory
$\alpha'(0)_{\rm pom}=0.25~$GeV$^{-2}$. This relation requires an input pair
$\sigma^T_0~$ , $B_0$ at a chosen energy. Using as input 
the $\sqrt{s}=541~{\hbox{\rm GeV}}$ data 
$\sigma^T_0=62.20$~mb, and $B_0=15.52~{\hbox {\rm GeV}}^{-2}~$
the line passes close to the CDF experimental point, instead of the 
E-710 point. If instead one uses as input $\sigma^T_0$ and 
$B_0$ from  the E-710 experiment at 
$\sqrt{s}=1800~{\hbox{\rm GeV}}$, the line shows a marked deviation 
at 541 GeV. This is rather intriguing, as it implies that 
the Regge formula favors the CDF experimental results at 1800 GeV. 
\bigskip

{\bf Acknowledgements}

Part of this work has been done while one of the authors (EF) was visiting 
CERN, and he wishes to thank the Theory Division for the hospitality.
Both authors thank CNPq (Brazil) and FAPERJ (Rio de Janeiro, Brasil) 
for financial support. The authors are grateful to H. G. Dosch for
discussions and criticism.

\newpage

\centerline{ CAPTIONS FOR THE FIGURES}
\bigskip


       Fig.1~-~Geometrical variables of the transverse plane, which
       enter in the calculation of the eikonal function
       for meson--meson scattering. The points $C_1$ and $C_2$ are the
       meson centres. In the integration, $P_2$ runs
       along the vector $\vec Q(2,1)$, changing
       the length $z$, which is the argument of the characteristic 
       correlator function. In analogous terms, points  $P_1$,
       $\bar P_1$ and $\bar P_2$ run along $\vec Q(1,1)$, $\vec Q(1,2)$ and
       $\vec Q(2,2)$. This explains the four terms that appear inside  the 
      brackets multiplying $\kappa$ in the expression for the loop-loop 
      amplitude. The length $z'$ of 
      the dot-dashed line is the argument of the Bessel function arising 
       from the non-confining correlator $D_1$; there are four such terms,
       appearing inside the brackets multiplying $(1-\kappa)$ . 
\bigskip
   
 Fig.2~-~ Relation between the two experimental quantities of 
the pp and $\bar{\rm p}{\rm p}$ systems. 
The values of $\sigma^T$ at energies up to 62.3 GeV shown in this figure are 
the $\sigma^T_{\rm pom}$  values as given by the parametrization  
 $\sigma^T_{\rm pom}=(21.70~{\rm mb}) s^{0.0808}$ of Donnachie-Landshoff.  
The values taken for B at the ISR energies are those for the pp system.
The solid line represents $ B=B_\Delta + C_\Delta  (\sigma^T)^\Delta$ ,
 with values for $\Delta$, $B_\Delta$ and $C_\Delta$ obtained by fitting 
the data. The point at 1800 GeV used in the determination of the line is 
taken from the Fermilab E-710 experiment, but the CDF point is also shown. 
\bigskip

   
  Fig.3~-~Constraints on the values of $\langle g^2 FF\rangle$ and of the
 correlation length $a$. The solid line is the fit of our correlator to the 
lattice  calculation \protect \cite{RRF1} as given in  
 eq.(\protect \ref{5A38}). The 
 dashed line plots  eq.(\protect \ref{5A8}), with  $\rho=0.16 $ GeV$^2$. 
 The cross centered at $a=0.32$ fm, $<g^2 FF>$=2.7 GeV$^{4}$ shows our 
  results.  
\bigskip

 
   Fig.4~-~ Energy dependence of the proton radius. The marked points are
obtained from the total cross-section data (at the ISR energies the total 
cross-sections are represented by the pomeron exchange contributions).
The two representations for the radius dependence are indistinguishable
with the present data, but give quite different predictions for the 
cross-section values at the LHC energies.  
\bigskip
\newpage

\bigskip
\centerline{TABLE AND TABLE CAPTION}
\bigskip
   Table 1. Values of the parameters for eq.(\ref{4A27}) 
  for  $\kappa=3/4$ and  $\kappa=1$  . 
\bigskip
 \begin{center}
\begin{tabular}{|c|c|c|c|c|c|c|}
  \hline
  $\kappa$& $\alpha\times 10^2 $ & $\beta$ & $\eta $ & $\gamma $& $\delta$ &
     $\delta / \beta$  \\
 \hline
 3/4&0.6532&2.791&2.030&0.3293&2.126&0.762\\
 \hline
 1.0 & 0.6717 &  3.029  & 1.859 & 0.3118 &  2.183 & 0.721  \\
 \hline
\end{tabular}
  \end{center}

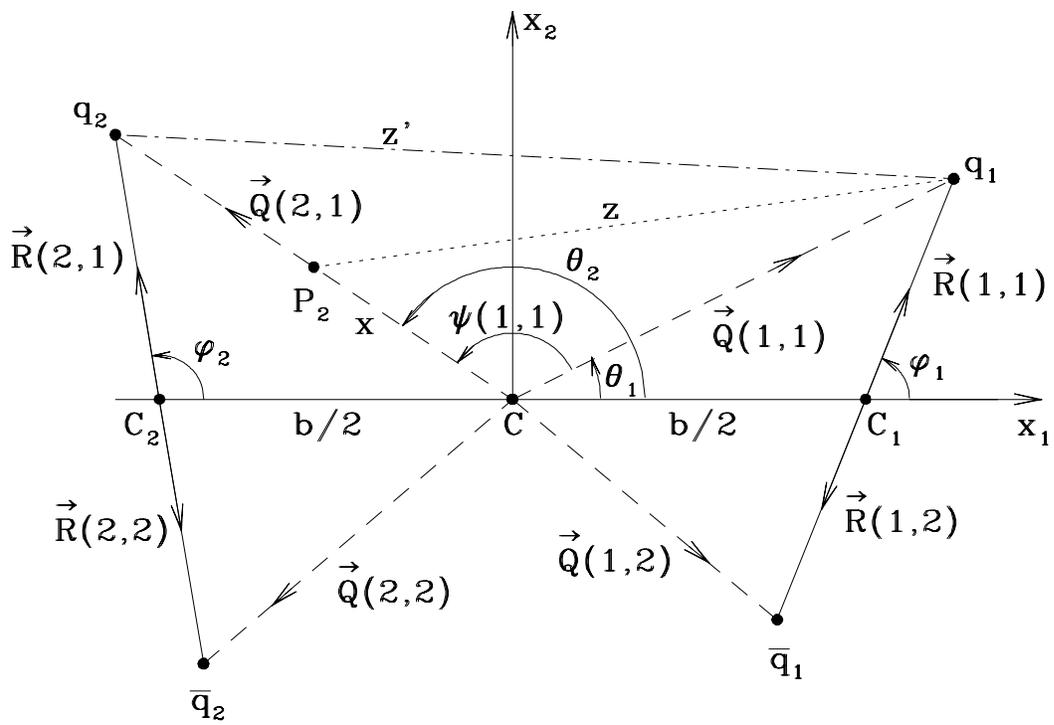
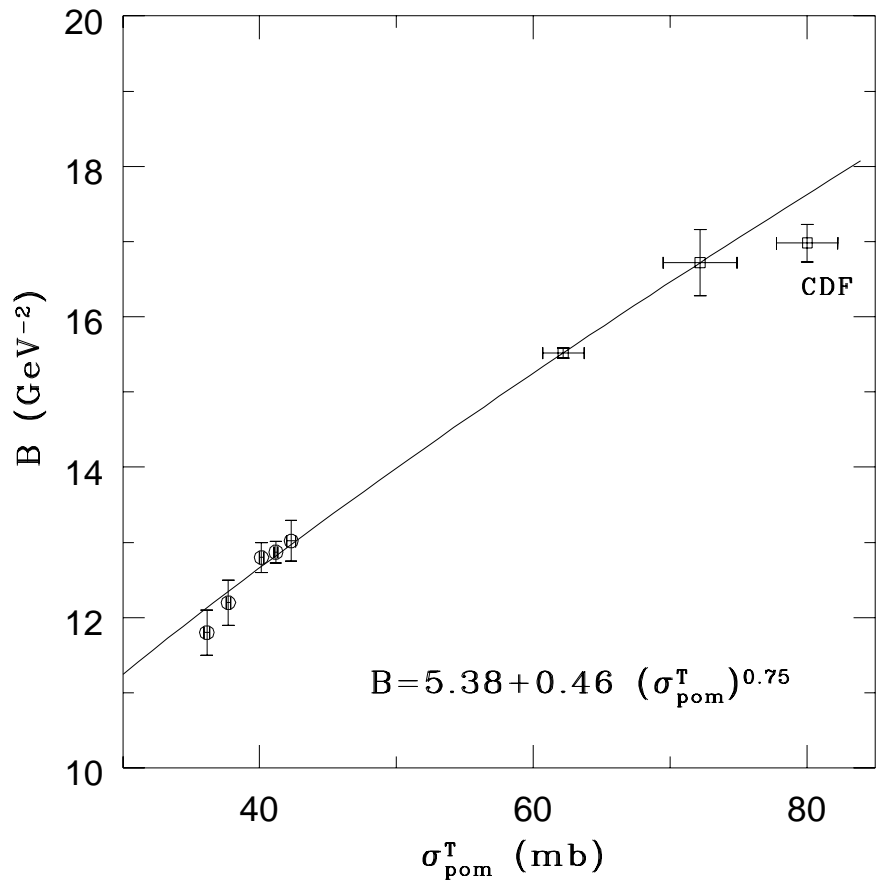
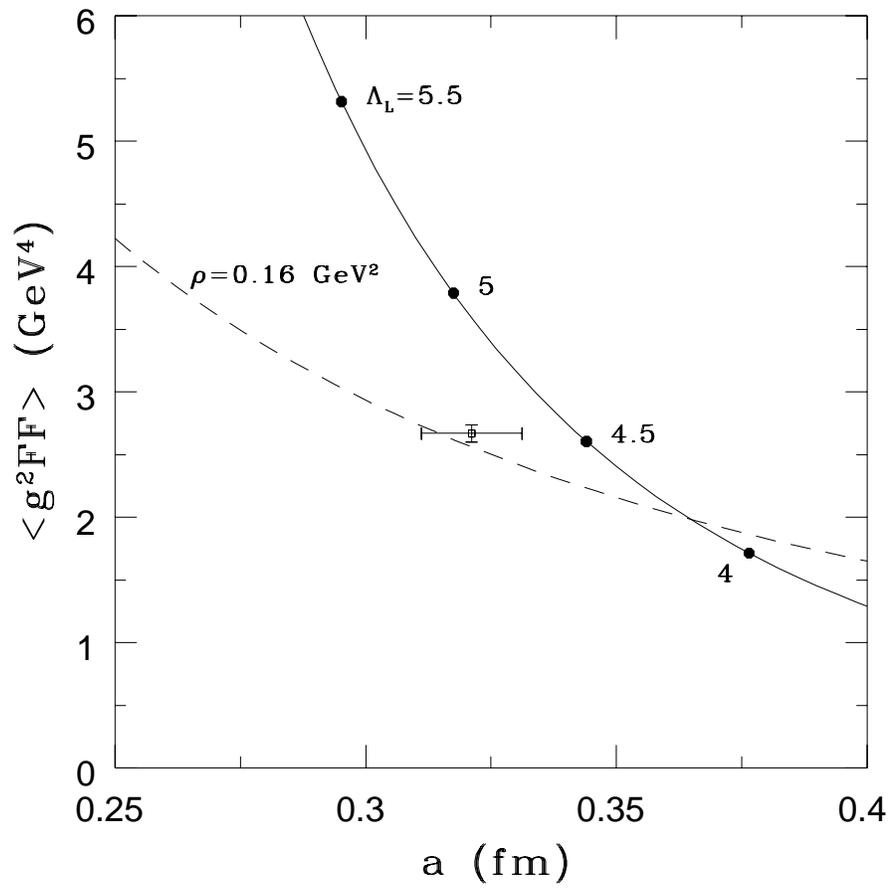
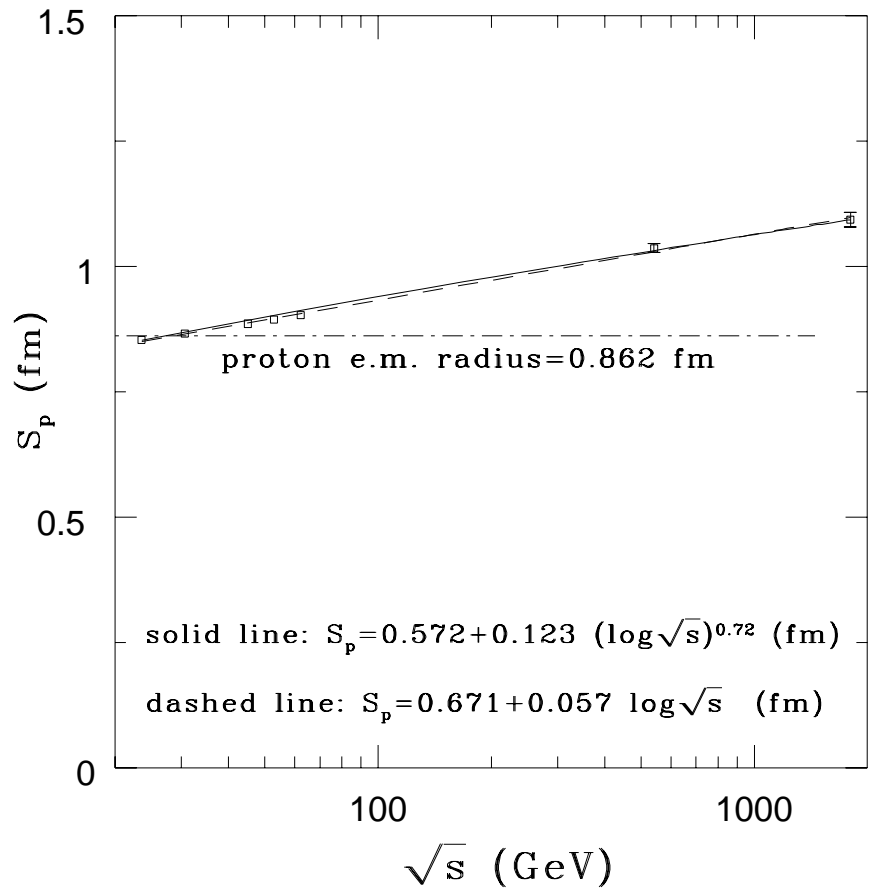
\end{document}